# On the emission probability of the 66.7 keV $\gamma$-transition in the decay of $^{171}$Tm

I. Kajan[1], S. Heinitz[1], R. Dressler[1], P. Reichel[1], N. Kivel[1], D. Schumann[1]

[1] Paul Scherrer Institute, 5232, Villigen PSI, Switzerland

*Devoted to Peder Gregers Hansen (1933 – 2005)*

Abstract: The $\gamma$-emission probability of the 66.73 keV line in the decay of $^{171}$Tm has been experimentally determined using $\gamma$-spectrometry and inductively coupled mass spectrometry. Using a set of two reference sources, namely $^{60}$Co as primary standard and $^{44}$Ti/$^{44}$Sc in secular equilibrium as secondary standards, we were able to deduce the detection efficiency at 67.87 keV of the used $\gamma$-spectrometry setup with high precision. The emission probability of the 66.73 keV $\gamma$-transition in the decay of $^{171}$Tm has been determined with a set of three radioisotopically pure $^{171}$Tm samples to be 0.159(6) %.

## Introduction

The radioactive isotope $^{171}$Tm, $t_{1/2}$ = 1.92(1) a, is a $\beta$-emitter with a low Q-value of 96.4(10) keV [1]. This isotope represents one out of 21 astrophysical interesting branching point nuclei along the s-process path [2]. The competition between the $\beta$-decay of $^{171}$Tm and an additional neutron capture influences the abundance of stable $^{171}$Yb and $^{172}$Yb. Recently, the isotopic ratio of these isotopes was measured in pre-solar SiC grains showing significant deviations from stellar models describing the evolution of asymptotic giant branch stars [3]. This discrepancy might be caused by insufficient knowledge of the nuclear properties of $^{171}$Tm. Very recently experiments were performed to investigate the neutron capture cross-section of $^{171}$Tm [4] but precise $\gamma$-spectroscopic characterization [5] of the amount of $^{171}$Tm target material used in these experiments strongly depends on the $\gamma$-emission probability of this isotope.

$^{171}$Tm can be produced by neutron activation of $^{170}$Er and is primarily measured by liquid scintillation counting. However, it also exhibits a weak $\gamma$-transition at 66.73 keV, which can be used to deduce its activity by $\gamma$-spectroscopy using HPGe detectors. The first estimate of emission probability $I_\gamma$ of this $\gamma$-line, with value $I_{66.7}$ = 0.14 %, given with no uncertainty, dates back to 1964 [6]. Within the work of P.G. Hansen, $^{171}$Tm has been produced by neutron irradiation of $^{170}$Er and was separated by ion exchange chromatography. The sample was then measured with a NaI scintillation detector, a Xe-filled proportional counter and an x-ray escape spectrometer. The internal-conversion line spectrum and the $\beta$ spectrum were additionally measured with a six-gap spectrometer. The $\gamma$-emission probability was then calculated based on assumed probabilities of K

and L shell electron transitions during the $^{171}$Tm decay. Very recently, the 66.73 keV $\gamma$-line intensity has been measured using neutron activation of enriched $^{170}$Er [7].

Within this work, the number of $^{171}$Tm atoms was determined by following the decay of $^{171}$Er, subsequently determining the 66.73 keV photon emission rate using $\gamma$-spectrometry. The knowledge of the exact emission probability and its uncertainty plays a crucial role in investigations involving this isotope, where the quantification of $^{171}$Tm is based on $\gamma$-spectrometry [5,8-9]. Complementary to parallel efforts described in [7], we report here on the re-determination of $I_{66.7}$ providing an uncertainty analysis following the recommendations of the "guide to the expression of uncertainty in measurement" (GUM) [10]. The determination of the emission probability of the 66.73 keV $\gamma$-line of $^{171}$Tm was based on measuring the number of $\gamma$ photons emitted from a quantified mass of $^{171}$Tm. This task was accomplished by using a combination of $\gamma$-spectrometry and mass spectrometry with radioisotopically pure solutions of $^{171}$Tm.

All uncertainties are calculated according to GUM [10] and are quoted with a coverage factor k = 1, i.e. confidence level of about 68 %, if not stated otherwise. The combined standard uncertainties are given in brackets in units of last significant digit.

## Experimental

Radioisotopically pure $^{171}$Tm was taken from leftovers of an earlier experiment performed in 2014 [4], where 140 GBq of $^{171}$Tm were produced and separated from 250 mg of 98.1 % enriched $^{170}$Er [5]. The last separation date of Tm from Yb was the 01.11.2014 with a total content of $^{171}$Yb of only 0.017 % [5]. The specific activity of the $^{171}$Tm solution was approx. 200 MBq/mL as of August 2017. An aliquot of 20 μL of this solution was diluted with 1 mL of MilliQ water (resistivity 18.2 MΩ cm at 25 °C) provided by an in-house water purification system. Three different $^{171}$Tm samples were prepared from this solution by evaporating one droplet (20 μL) on circular Teflon supports. A Mettler-Toledo AT261 DeltaRange (d = 0.01 mg) balance was used to gravimetrically trace all dilution and sample preparation steps. The total weight contribution of the 171 isobar, i.e. the abundance of ($^{171}$Tm + $^{171}$Yb) in the used solution, was measured 8 times by HR-ICP-MS using an Element 2, Thermo-Fischer Scientific, Bremen, Germany, operated in low resolution mode and wet plasma conditions.

A secondary calibration source was prepared from a purified stock solution $^{44}$Ti/$^{44}$Sc available from earlier experiments [11]. An aliquot of this stock solution was diluted with MilliQ water to yield a solution containing approx. 50 kBq/mL of $^{44}$Ti/$^{44}$Sc. From this dilution, a $^{44}$Ti source was prepared by evaporating 20 μL of the solution on a Teflon support. This source and the $^{171}$Tm samples were sealed using a 50 μm thick Kapton tape.

A $^{60}$Co calibration source served as primary reference standard with an activity of 1831(13) Bq certified with k = 2 on 01.07.2011 by Physikalisch-Technische-Bundesanstalt (PTB), Braunschweig, Germany. All sources and the $^{171}$Tm samples were measured at a distance of 140(1) mm of air on a 1.2 mm thick Al support in front of a Canberra BE2825 planar HPGe detector for at least 21 h to achieve not less than $6.5 \times 10^4$ full energy peak (FEP) net counts in the regions of interest. The primary standard ($^{60}$Co) was measured five consecutive times, while the $^{44}$Ti/$^{44}$Sc source and each $^{171}$Tm sample were measured three times. The dead time of the detector was always below 0.5 %. The energy calibration of the detector was performed by a mixed nuclide standard provided by the Czech metrology institute, Brno, Czech Republic, containing $^{241}$Am, $^{109}$Cd, $^{139}$Ce, $^{57}$Co, $^{113}$Sn, $^{85}$Sr, $^{137}$Cs, $^{60}$Co and $^{88}$Y with certified activities. True coincidence summing effects for the measured geometry at 140(1) mm distance were calculated utilizing the EFFTRAN code [12]. True coincidence summing correction factors (ranging between: 1.004 - 1.009) were taken into account during the data evaluation process for all measured nuclides. Background measurements revealed no interfering peaks in the regions of interest. Due to the close proximity of the $\gamma$-line of $^{171}$Tm to the ones of $^{44}$Ti, no correction for attenuation effects in air, Aluminium or Teflon was performed. Attenuation effects of the high-energy lines of $^{44}$Sc and $^{60}$Co were considered as negligible. Values for efficiencies as well as the $\gamma$-line emission probability of $^{171}$Tm were obtained from the weighted average of replicated measurements.

## Results

The measured $\gamma$-spectra of the $^{60}$Co reference source, the $^{44}$Ti/$^{44}$Sc and $^{171}$Tm samples are given in Figure 1.

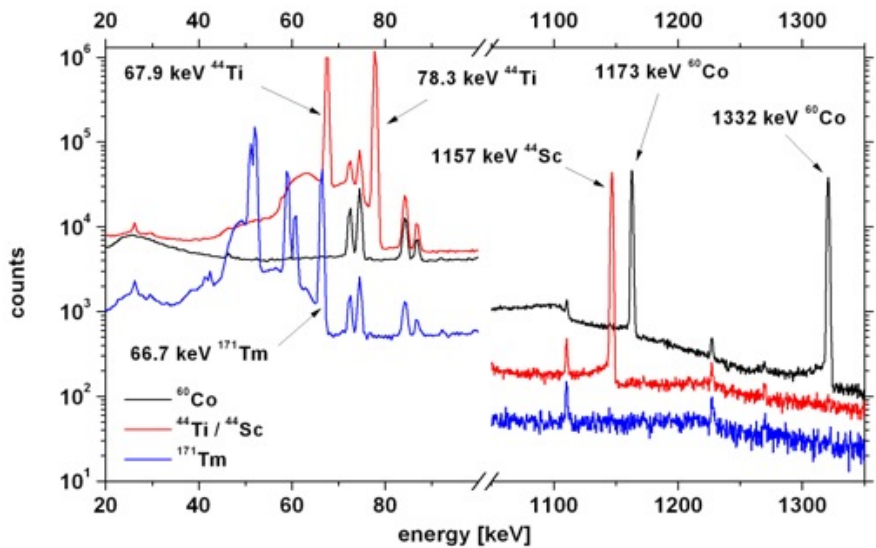

Figure 1: Superimposed $\gamma$-spectra of the $^{60}$Co, $^{44}$Ti/$^{44}$Sc and $^{171}$Tm samples.

Applying decay correction with respect to the center of the counting interval for the measured primary $^{60}$Co calibration source, the (FEP) efficiencies at 1173.23 keV and 1332.49 keV ($I_{1173}$ = 99.85(3) % and $I_{1332}$ = 99.9826(6) % [13]) were determined to be $\varepsilon_{1173}$ = 0.0666(3) % and $\varepsilon_{1332}$ = 0.0587(2) %, respectively, for the measured geometry.

The FEP efficiency $\varepsilon_x$ and its relative uncertainty $\Delta\varepsilon_x/\varepsilon_x$ of a given detection system at an energy $E_x$ is extrapolated from two known FEP efficiencies $\varepsilon_1$ and $\varepsilon_2$ measured at energies $E_1$ and $E_2$, where $E_x < E_1 < E_2$, assuming a general power-law using the following equations 1 and 2[1]:

$$\varepsilon_x = \varepsilon_1 \cdot \left(\frac{E_x}{E_1}\right)^\alpha = \varepsilon_1 \cdot \left(\frac{\varepsilon_2}{\varepsilon_1}\right)^\beta \tag{1}$$

$$\left(\frac{\Delta\varepsilon_x}{\varepsilon_x}\right)^2 = \left(\frac{\Delta\varepsilon_1}{\varepsilon_1}\right)^2 + \beta^2 \cdot \left(\left(\frac{\Delta\varepsilon_1}{\varepsilon_1}\right)^2 + \left(\frac{\Delta\varepsilon_2}{\varepsilon_2}\right)^2\right) \tag{2}$$

where $\alpha$ and $\beta$ are defined as

$$\alpha = \frac{\ln\left(\frac{\varepsilon_2}{\varepsilon_1}\right)}{\ln\left(\frac{E_2}{E_1}\right)} \qquad \beta = \frac{\ln\left(\frac{E_x}{E_1}\right)}{\ln\left(\frac{E_2}{E_1}\right)}.$$

With the above equations, the FEP efficiency at 1157.02 keV was determined using the FEP efficiencies of both $^{60}$Co lines to be $\varepsilon_{1157}$ = 0.0676(3) %. Subsequently, the activity of $^{44}$Sc was determined by measuring the absolute count rate of its 1157.02 keV $\gamma$-line ($I_{1157}$ = 99.875(3) % [15]) to be 1287(7) Bq on reference date 22.09.2017. Due to secular equilibrium between $^{44}$Ti and its decay product $^{44}$Sc, the activities of both isotopes are equal.

In the next step, the FEP efficiencies of the $\gamma$-detector at both 67.87 keV ($I_{67.9}$ = 93.0(20) % [15]) and 78.32 keV ($I_{78.3}$ = 96.4(17) % [15]) $\gamma$-lines of $^{44}$Ti were determined using the secondary $^{44}$Ti standard. The FEP efficiency at these energies were $\varepsilon_{67.9}$ = 0.777(17) % and $\varepsilon_{78.3}$ = 0.784(14) %, respectively. In analogy to the extrapolation of the efficiency at 1157 keV, the FEP efficiency at 66.73 keV was determined to be $\varepsilon_{66.7}$ = 0.776(17) % by using equations 1 and 2.

Finally, measurements of $^{171}$Tm revealed the absolute $\gamma$-emission rate of the 66.73 keV line of $^{171}$Tm emitted from each prepared sample. Measured data from three different $^{171}$Tm samples standardized to the reference date 22.9.2017 are given in Table 1.

Table 1: Measured characteristics of all three $^{171}$Tm samples used within this work; reference date 22.09.2017.

| Sample | Mass of ($^{171}$Tm+$^{171}$Yb) on sample [ng][a] | Specific activity of $^{171}$Tm [kBq] at reference date[b] | 66.73 keV photon emission rate [s$^{-1}$][c] |
|---|---|---|---|
| #1 | 5.37(15) | 76.2(22) | 0.941(3) |
| #2 | 5.53(16) | 78.4(23) | 0.969(3) |
| #3 | 5.26(15) | 74.6(22) | 0.923(3) |

[a] *Calculated from ICP-MS and gravimetrical measurements.*
[b] *Calculated according to decay corrections.*
[c] *Average originating from triplicated gamma spectroscopy measurements of each sample.*

---

[1] This procedure uses a power-law regression between two reference energies and is applied for extrapolations to energies lying in the vicinity of reference data points. For the present work the extrapolated energies are within a margin of 2% off from the closest reference energy and the approximations as well as their uncertainty are thus considered as acceptable. Refer to [14] for more details about the physical justification of this approach.

The concentration of the 171 isobar isotopes i.e. the content of $^{171}$Tm and $^{171}$Yb was measured by ICP-MS to be 260.3(73) ng/g in the prepared solution. With the known amounts of evaporated solution used to prepare the $^{171}$Tm sources, the last Tm/Yb separation date 01.11.2014 and the $^{171}$Tm half-life, we compute the 66.73 keV emission probability to be

$$I_{66.7} = 0.159(6) \%.$$

The contributors to the uncertainty of this value are shown in Table 2. An uncertainty propagation based on the values stated in Table 2. was made. The main contributor to systematic uncertainty originated from ICP-MS measurements as counting uncertainties from $\gamma$-measurements were within the range of 0.4 %. Other systematic uncertainties originate from literature data on emission probabilities of $\gamma$-lines used within the experiments as well as uncertainties attributed to half-life values of utilized radionuclides. The uncertainty originating from an incomplete separation Tm/Yb was evaluated to be below 0.1 %.

Table 2: Uncertainty contributions for the calculation of the emission probability

| Uncertainty source | Relative uncertainty [%] |
|---|---|
| $^{60}$Co calibration source | 0.36 |
| $^{60}$Co half-life | 0.007 |
| 1173 keV $\gamma$-emission probability | 0.03 |
| 1332 keV $\gamma$-emission probability | 0.001 |
| Counting statistics | < 0.36 |
| | |
| FEP efficiency at 1157 keV | 0.39 |
| 1157 keV $\gamma$-emission probability | 0.003 |
| 78.3 keV $\gamma$-emission probability | 2.1 |
| 67.9 keV $\gamma$-emission probability | 1.7 |
| Counting statistics | < 0.41 |
| | |
| FEP efficiency at 66.7 keV | 1.7 |
| $^{171}$Tm half life | 0.71 |
| $^{171}$Tm concentration | 2.8 |
| | |
| **Overall uncertainty** | **3.7** |

## Conclusions

The photon emission probability of 66.73 keV $\gamma$-line of $^{171}$Tm was measured by means of $\gamma$-spectrometry combined with ICP-MS measurements of radioisotopicaly pure $^{171}$Tm samples. The uncertainty of the measured value was determined using standard uncertainty propagation following the recommendations of GUM, whereas the main uncertainty contributor was originating from the ICP-MS measurement.

In comparison with the previously performed determinations of this quantity with a value of $I_{66.7}$ = 0.14 % by P.G. Hansen [6] and $I_{66.7}$ = 0.144(10) % by [7], our result of $I_{66.7}$ = 0.159(6) % is significantly higher. Our value is by 3$\sigma$ away from both previously measured $^{171}$Tm emission probabilities.

This new result and result from work of Weigand et.al.[7] leads to the weighted average value of $I_{66.7}$ = 0.155(5) %. This value will consequently reduce the deduced number of atoms in recently performed neutron capture experiments [4] thus enhancing the measured cross-sections by approx. 10 %.